\documentclass[pra,twocolumn,showpacs]{revtex4}

\usepackage{amsmath,amsfonts,amssymb,graphicx}

\begin{document}

\title{Inequivalent classes of closed three-level systems}

\author{Andrei B. Klimov}
\affiliation{Departamento de F\'{\i}sica,
Universidad de Guadalajara,
Revoluci\'on 1500, 44420~Guadalajara,
Jalisco, Mexico}

\author{Hubert de Guise}
\affiliation{Department of Physics,
Lakehead University, Thunder Bay,
Ontario, P7B 5E1 Canada}

\author{Luis L. S\'anchez-Soto}
\affiliation{Departamento de \'{O}ptica,
Facultad de F\'{\i}sica,
Universidad Complutense,
28040 Madrid, Spain}

\date{\today}

\begin{abstract}
We show here that the  $\Lambda$ and V
configurations of three-level atomic
systems, while they have recently been shown
to be equivalent for many important physical
quantities when driven with classical fields
[M. B. Plenio, Phys. Rev. A \textbf{62},
015802 (2000)], are no longer equivalent when
coupled via a quantum field. We analyze the
physical origin of such behavior and show
how the equivalence between these two configurations
emerges in the semiclassical limit.
\end{abstract}


\maketitle

Atomic coherence is essential to properly understand
the response of an atomic three-level system to
laser radiation (for a recent review see
Ref.~\cite{special}). A large amount of research
has thus been devoted to exploring many effects
that rely on quantum interference in atomic systems:
examples include dark states~\cite{Orriols76},
narrow spectral lines~\cite{narrow}, pulse
matching~\cite{Cook79}, and anti-intuitive
excitation~\cite{anti}. These nonclassical
features have an enormous variety of
interesting and nontrivial consequences,
including electromagnetically induced
transparency~\cite{EIT}, lasing without
inversion~\cite{LWI}, state-selective molecular
excitation~\cite{STIRAP}, and demonstrations of
slow light~\cite{slowlight} and fast
light~\cite{fastlight}, to mention only a few
examples.

Roughly speaking, one can identify dark states
as a key concept in the description of these
coherent phenomena: a dark state is a specific
coherent superposition resulting, by destructive
quantum interference, in a completely decoupled
state. So, the atom prepared in a dark state
cannot be excited and cannot leave the dark state.

When we limit the discussion to the case in
which only two transitions are allowed between
levels, there are three distinct level configurations
known as $\Xi$, $\Lambda$, and V~\cite{Yoo85}.
It is well known that dark states cannot be formed
in the $\Xi$ configuration: this is the reason
why quantum interference does not play any
role for this system and it is tacitly
considered inequivalent to $\Lambda$ and
V configurations.

Usually the phenomenon of spontaneous emission
(which is the main damping contribution) plays a
destructive role in the creation of this coherence.
Note, however, that there have been several
proposals in which coherence induced from the
spontaneous emission  itself is used for the
preparation of the atom~\cite{spont}.

One can find, scattered in the literature,
statements about similarities between
$\Lambda$- and V-type systems in some limits
or under different decaying rates between
levels~\cite{equivlim}. The recent and
intriguing paper by Plenio~\cite{Ple00}
sheds light on these similarities, pointing
out a more general equivalence between these
systems. The central result is that both schemes
share a common structure and, as a consequence,
exhibit the same physical behavior for many
important quantities. In general, to derive
this equivalence, one writes the master
equations of both systems and, by a smart
change of variables, shows that these
equations are identical.

Although such equivalence between $\Lambda$
and V schemes is valid for many purposes,
it is always deduced in a regime where the
fields are essentially classical. It remains
to investigate the extent to which these systems
remain equivalent when they interact with quantum
fields. It is precisely the objective of this
paper to answer this question.

We begin by considering a collection of $A$
identical three-level atoms confined to a
small volume with linear dimensions less
than the relevant wavelengths of light.
The atomic energy levels are always ordered
according $E_1 \le E_2 \le E_3$. The
collective atomic operators are denoted
by $S_{ij}$ (the latin indices run from 1
to 3) and satisfy the commutation
relations
\begin{equation}
[ S_{ij}, S_{kl}]=
\delta _{jk} S_{i l} -
\delta_{i l} S_{kj} \, ,
\end{equation}
distinctive of the algebra u(3). For
concreteness, we shall treat only fully
symmetrical states; then, $S_{ij}$ is
conveniently realized, in the second
quantization formalism, by boson operators
\begin{equation}
\label{boson}
S_{ij} = b_i^\dagger b_j \, ,
\end{equation}
which transfer excitations from level $j$ to
level $i$ $(i \ne j)$. The eigenvalue of
$S_{ii}$ is just the population of level $i$.

The atomic system interacts with a single
quantum field of frequency $\omega$ described
by the usual creation and destruction operators
$a^\dagger$ and $a$ (the case of two-mode
fields  can be treated much in the same way).
The general form of the Hamiltonian for our
systems is $H = H_0 + H_{\mathrm{int}}$,
where the free Hamiltonian is
(in units $\hbar = 1$)
\begin{equation}
H_0 = \sum_{i=1}^{3}\, E_i S_{ii} +
\omega \, a^\dagger a \, ,
\end{equation}
and the interaction Hamiltonian depends
on the level configurations:
\begin{eqnarray}
H^{(\Lambda)}_{\mathrm{int}} & = &
g_{31} ( X_{31} + X^\dagger_{31} ) +
g_{32} ( X_{32} + X^{\dagger}_{32} ) \, ,
\nonumber \\
& & \\
H^{(\mathrm{V})}_{\mathrm{int}} & = &
g_{31} ( X_{31} + X^\dagger_{31} ) +
g_{21} ( X_{21} + X^{\dagger}_{21} ) \, ,
\nonumber
\end{eqnarray}
which have been written in terms of
operators in the so-called su(3) deformed
algebra~\cite{defalg}
\begin{equation}
\label{deformat}
X_{31} =  a S_{31} \, ,
\quad
X_{21} = a S_{21}  \, ,
\quad
X_{32} = a S_{32}  \, ,
\end{equation}
and $X_{ij} = (X_{ji})^\dagger$. These first-order
transition operators describe allowed (direct)
transitions between the corresponding atomic
levels, accompanied by the appropriate emission
or absorption of a photon.

Note that a pair of levels (lower levels, in
the case of $\Lambda$, upper levels, in the
case of V) must be nearly degenerate in order to
interact efficiently with a single field mode.
In the case of degenerate levels for
the $\Lambda$ system, we write $E_1 = E_2
\equiv E_-$, $E_3 \equiv E_+$, and rotate
the operators $b_j$, which enter in the
representation of atomic operators (\ref{boson}),
to
\begin{equation}
\left(
\begin{array}{c}
b_1 \\
b_2
\end{array}
\right) =
\left (
\begin{array}{cc}
\cos \alpha & -\sin \alpha \\
\sin \alpha & \cos \alpha
\end{array}
\right )
\left (
\begin{array}{c}
c_1 \\
c_2
\end{array}
\right ) ,
\qquad
b_3 = c_3 \, ,
\label{transfo}
\end{equation}
where $c_1$ and $c_2$ are new destruction operators
and $\tan \alpha = g_{32}/g_{31}$. In terms of new
atomic operators $\tilde{S}_{jk}= c_j^\dagger c_k$
the transformed Hamiltonian becomes
\begin{eqnarray}
\label{hLambda}
\tilde{H}^{(\Lambda)} & = & h^{(\Lambda)} +
E_- \tilde{S}_{22} \, , \nonumber \\
&  & \\
h^{(\Lambda)}  & = &  \omega \, a^\dagger a +
E_+ \tilde{S}_{33} + E_- \tilde{S}_{11} +
g_{\Lambda}(a \tilde{S}_{31}+a^\dagger
\tilde{S}_{13}) \, , \nonumber
\end{eqnarray}
where $g_{\Lambda} = g_{31} \cos \alpha +
g_{32} \sin \alpha$.

The dynamics of the uncoupled level
$| \tilde{2} \rangle = -\sin \alpha
| 1 \rangle + \cos \alpha | 2 \rangle $ is
completely independent of the field variables
and is governed by the subHamiltonian $E_-
\tilde{S}_{22}$ of $\tilde{H}^{(\Lambda)}$.
The levels $| \tilde{1} \rangle $ and
$| \tilde{3} \rangle $ are coupled via
an effective coupling constant $g_{\Lambda}$.

The same procedure can be repeated for
$H^{(\mathrm{V})}$, rotating this time
$b_2$ and $b_3$. Using $E_1 = E_-$
and $E_2 = E_3 \equiv E_+$, we obtain
\begin{eqnarray}
\label{hvee}
\tilde{H}^{(\mathrm{V})} & = & h^{(\mathrm{V})} +
E_+ \tilde{S}_{22} \, , \nonumber  \\
& & \\
h^{(\mathrm{V})} & = &  \omega a^\dagger a +
E_- \tilde{S}_{11}+ E_+ \tilde{S}_{33}+
g_{\mathrm{V}} (a \tilde{S}_{31} + a^\dagger
\tilde{S}_{13}) \, ,
\nonumber
\end{eqnarray}
where $g_{\mathrm{V}} = g_{21} \sin
\beta + g_{31} \cos \beta$ and $\tan \beta =
g_{21}/g_{32}$. A simple look at the
transformed Hamiltonians (\ref{hLambda})
and (\ref{hvee}) immediately shows that
they both have dark states and the
dynamics of the remaining two-level
subsystems is the same. This is the
basis on which rests the claim of
dynamical equivalence between $\Lambda$
and V configurations.

On closer examination, the complete equivalence
between $\Lambda$ and V configurations should
also include higher-order processes, since
the action of first-order operators defined
in (\ref{deformat}) is intrinsically
nonlinear~\cite{defalg}. In what follows,
we concentrate on second-order feasible
processes like the one represented by $X_{23}
X_{31} = a^\dagger a ( S_{33} + 1 ) S_{21}$
(for the $\Lambda$ scheme). This results in
a net transfer of one atomic excitation
between the degenerate levels  $| 1 \rangle $
to $| 2 \rangle$, with the transition enhanced
by a factor which depends both on field and
atomic populations.  This enhancement is
different for the process $X_{31}X_{23} =
(a^\dagger a + 1 ) S_{33} S_{21},$ which also
results in a transfer from $| 1 \rangle $ to
$| 2 \rangle$, so that the second-order
operator
\begin{equation}
\label{lambda}
[ X_{31}, X_{23}] =
( S_{33}- a^\dagger a ) S_{21} \, ,
\end{equation}
describes an effective intensity-dependent
transition $1 \leftrightarrow 2$, stimulated
by the field strength and the population in the
intermediate level. In particular, the
commutator will vanish if the population
of the intermediate state is precisely
equal to the total number of photons
in the system.

Take now the second-order operator for the
V scheme
\begin{equation}
\label{v}
[ X_{31}, X_{12} ] = ( S_{11} +
a^\dagger a + 1 ) S_{32} \, ,
\end{equation}
which measures the difference between two
two-step transfers of excitation between the
degenerate upper atomic levels. Again, this
second-order operator depends on the population
of the intermediate state and on the photon
population. However, this commutator will never
vanish.

\begin{figure}[h]
\centering
\resizebox{0.85\columnwidth}{!}{\includegraphics{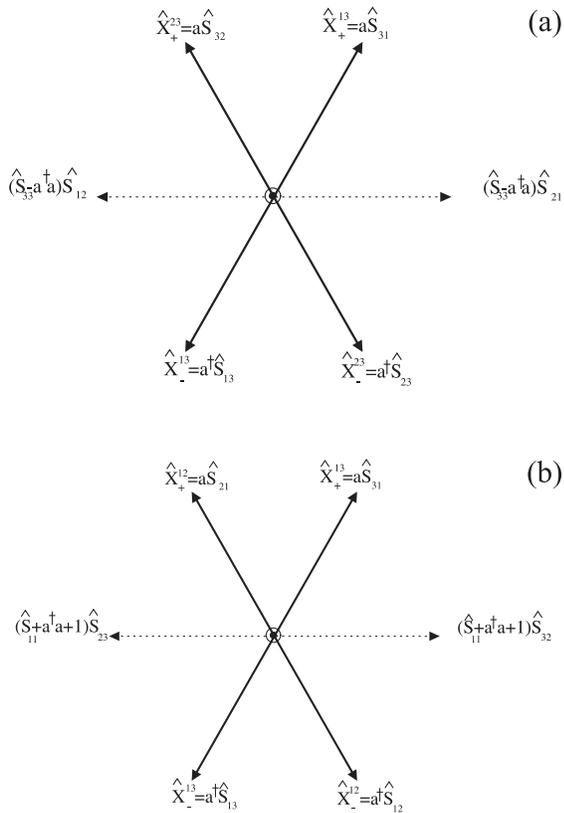}}
\caption{A root-like diagram for the first-order
operators (thick lines) and second order operators
(dashed lines) for (a) $\Lambda$  and (b)  V
schemes.}
\label{fig1}
\end{figure}

To graphically illustrate the differences between
both schemes, we shall use a root--like diagram
constructed in the following way. We choose a
Cartan subalgebra (i. e., maximal set of commuting
operators) containing the two independent inversions
$h_1= S_{11}- S_{22}$ and $ h_2 =  S_{22}- S_{33}$
for $\Lambda$ and $h_1= S_{22}- S_{11}$ and
$h_2 = S_{33}- S_{22}$ for V. Then, we define the
weight components $\kappa_1$ and $\kappa_2$ through
the ``eigenvalue" equations for first-order operators
in (\ref{deformat}),
\begin{equation}
[ h_1, X_{ij} ] = \kappa_1 X_{ij} ,
\qquad
[ h_2, X_{ij} ] = \kappa_2 X_{ij} \, ,
\end{equation}
and analogously for second-order operators.
The eigenvalues $(\kappa_1, \kappa_2)$ obtained for
relevant first- and second-order operators are then
placed on a two-dimensional diagram, using as basis
the vectors of the su(3) root diagram, which are
angled at $2\pi/3$ to one another. One then draws
from the center vectors to the points on the diagram.
This is illustrated in Fig.~\ref{fig1}.

We recall that the major feature of this weight
diagram is that commutation is mapped, up to a
multiplicative factor, to vector addition~\cite{Bar77}.
For instance, the result of $[X_{31}, X_{23} ] $
is proportional to the vector resulting from
the addition of the root vectors for $X_{31}$
and $X_{23}$.

It is clear by inspection of Eqs.~(\ref{lambda})
and (\ref{v}), and from Fig.~\ref{fig1}, that
$\Lambda$ and V configurations are not equivalent
when coupled via a quantum field: no unitary transformation
acting on the atomic operators can transform
$H^{(\Lambda)}$ into $H^{(\mathrm{V})}$. In particular,
no relabeling of the atomic states transforms a $
\Lambda$ into a V: the structure of the second-order
operators prevents this.

As one would expect, all differences vanish in systems
where first-order operators contain classical rather
than quantum fields. If $a$ is replaced by a
complex number $\alpha$, the transition operators
of both schemes reduce to su(3) operators and close
on equivalent su(3) algebras: in the $\Lambda$ case,
the first-order operators become $\alpha S_{32},
\alpha S_{31}$ and their conjugates while second-order
operators $[\alpha S_{31}, \alpha^\ast S_{23}]$ reduces
to $-\alpha^\ast \alpha S_{21}$. A similar argument
applies to the V case. The equivalence found by
Plenio can then simply be expressed by the statement
that the first-order operators for $\Lambda$ can be
transformed into the first-order operators for
$V$ by geometrical reflection of the root vector;
it is this reflection which effects the relabeling
of basis states proposed in Ref.~\cite {Ple00}.

Although differences will certainly be noticeable
when the number of field quanta and the level
population are both low, we observe that these
differences can be important even in strong
fields when the number of atoms $A$ is large.

As a simple though remarkable application of
the above discussion, we consider the dynamics
of the $\Lambda $ and V configurations in the
dispersive regime, when
\begin{equation}
|\Delta _{ij}| \gg A g_{ij}
\sqrt{ \langle a^\dagger a \rangle +1}\, ,
\end{equation}
with $ \Delta_{ij}= E_{i}- E_{j}-  \omega$.
Following Ref.~\cite{Kli00}, let us define the
following unitary tranformations:
\begin{equation}
U (\varepsilon_{ij})=  \exp [ \varepsilon_{ij}
(X_{ij} - X_{ij}^\dagger) ]\, ,
\end{equation}
where
\begin{equation}
\varepsilon_{ij}= \frac{g_{ij}}{\Delta _{ij}} \ll 1 \,,
\end{equation}
are small parameters in this regime.
One then shows that
$\tilde{H}_{\mathrm{eff}}^{(\Lambda)} =
U(\varepsilon_{32}) U(\varepsilon_{31})
H^{(\Lambda)} U^\dagger (\varepsilon_{31})
U^\dagger (\varepsilon_{32})$ is the effective
Hamiltonian
\begin{equation}
\label{HeffL}
\tilde{H}_{\mathrm{eff}}^{(\Lambda)}=
\varepsilon_{31}g_{32}
( S_{12} + S_{12}^\dagger ) (S_{33}- a^\dagger a) \, ,
\end{equation}
where we have omitted diagonal terms that contain
the dynamical Stark shift~\cite{Kli02}. It is clear
that there will be no population transfer
between levels $| 1 \rangle$ and $| 2 \rangle$
when the population of $| 3 \rangle$ is exactly
equal to the number of field quanta. In particular,
there will be no transfer if the field is in the vacuum
and level $| 3 \rangle$  is unoccupied.

Applying the same method to $H^{(\mathrm{V})}$, we get
$\tilde{H}_{\mathrm{eff}}^{(\mathrm{V})} =
U(\varepsilon_{31}) U(\varepsilon_{21})
H^{(\mathrm{V})} U^\dagger (\varepsilon_{21})
U^\dagger (\varepsilon_{31})$, where
\begin{equation}
\label{HeffV}
\tilde{H}_{\mathrm{eff}}^{(\mathrm{V})}=
\varepsilon_{21}g_{13}
( S_{32}+ S^\dagger_{32}) (S_{11} + a^\dagger a + 1) \,.
\end{equation}
In contrast with the results for the $\Lambda $
configuration, there is \textit{always} a population
transfer between the degenerate levels $| 2 \rangle$
and $| 3 \rangle$ via the intermediate level in the
V configuration. The transfer of excitations between
levels $| 2 \rangle$ and $| 3 \rangle$ in the V
configuration takes place even when level
$| 2 \rangle$ is unpopulated and there are no
field quanta. This occurs because of the spontaneous
emission (stimulated by the zero-point fluctuations
of the quantum field) from the $| 2 \rangle$ to
$| 1 \rangle$ with a subsequent absorption of the
emitted photon leading to the population of the
upper levels. Once again, these differences disappear
in the limit of classical fields.

In conclusion, we have shown that the $\Lambda$ and V
configurations cannot be taken as equivalent if we
treat the photon field as a quantized field. It is
also possible to see why, physically, these
configurations are different: in a V configuration,
vacuum fluctuations can create a photon, the
absorption of which acts as a trigger for
the transfer of excitation between levels
$| 2 \rangle$ and $| 3 \rangle$. This transfer
mechanism cannot occur in the $\Lambda$
configuration.

It is the hope that these basic results will
help to elucidate the origin of equivalences
between different three-level schemes,
also when  extra decay rates for the levels
are taken into account.

The work of HdG is supported by  NSERC of Canada.


\begin{thebibliography}{9}

\bibitem{special}
J. Mod. Opt. \textbf{49}, 1 (2002),
special issue on quantum interference.

\bibitem{Orriols76}
E. Arimondo and G. Orriols,
Lett. Nuovo Cim. \textbf{17}, 333 (1976);
G. Alzetta, A. Gozzini, L. Moi, and G. Orriols,
Nuovo Cim. B \textbf{36}, 5 (1976);
R. M. Whitley and C. R. Stroud,
Phys. Rev. A \textbf{14}, 1498 (1976);
E. Arimondo, Prog. Opt. \textbf{35}, 257 (1996).

\bibitem{Hegerfeldt92}
G. C. Hegerfeldt and M. B. Plenio,
Phys. Rev. A \textbf{46}, 373 (1992).

\bibitem{narrow}
G. C. Hegerfeldt and M. B. Plenio,
Phys. Rev. A \textbf{52}, 3333 (1995);
M. B. Plenio,
J. Mod. Optics \textbf{43}, 753 (1996);
P. Zhou and S. Swain,
Phys. Rev. Lett. \textbf{77}, 3995 (1996);
E. Paspalakis and P. L. Knight,
{\em ibid.} \textbf{81}, 293 (1998);
C. H. Keitel,
{\em ibid.} \textbf{83}, 1307 (1999).

\bibitem{Cook79}
R. J. Cook and B. W. Shore,
Phys. Rev. A \textbf{20}, 539 (1979);
F. T. Hioe and J. H. Eberly,
{\em ibid.} \textbf{25}, 2168 (1982);
S. E. Harris,
Phys. Rev. Lett. \textbf{72}, 52 (1994);
J. H. Eberly, M. L. Pons, and H. R. Haq,
{\em ibid.} \textbf{72}, 56 (1994).

\bibitem{anti}
J. Oreg, F. T. Hioe, and J. H. Eberly,
Phys. Rev. A \textbf{29}, 690 (1984).

\bibitem{EIT}
S. E. Harris, J. E. Field, and A. Imamouglu,
Phys. Rev. Lett. \textbf{64}, 1107 (1990);
K. J. Boller, A. Imamoglu, and S. E. Harris,
{\em ibid} \textbf{66}, 2593 (1991);
S. E. Harris,
{\em ibid.} \textbf{70}, 552 (1993);
A. Kasapi, M. Jain, G. Y. Yin, and S. E. Harris
{\em ibid.}\textbf{ 74} 2447 (1995);
S. E. Harris,
Physics Today \textbf{50}, 36 (1997);
E. Paspalakis, S. Q. Gong, and P. L. Knight,
Opt. Commun. \textbf{152}, 293 (1998);
E. Paspalakis, N. J. Kylstra, and P. L. Knight,
Phys. Rev. Lett. \textbf{82}, 2079 (1999).

\bibitem{LWI}
O. A. Kocharovskaya and Ya. I. Khanin,
JETP Lett. \textbf{48}, 580 (1988);
S. E. Harris,
Phys. Rev. Lett. \textbf{62}, 1033 (1989);
M. O. Scully, S. Y. Zhu, and A. Gavrielides,
{\em ibid.} \textbf{62}, 2813 (1989);
A. Imamoglu, J. E. Field, and S. E. Harris,
{\em ibid.} \textbf{66}, 1154 (1991);
G. G. Padmabandu, G. R. Welch, I. N. Shubin,
E. S. Fry, D. E. Nikonov, M. D. Lukin, and M. O. Scully,
{\em ibid.} \textbf{76}, 2053 (1996).

\bibitem{STIRAP}
U. Gaubatz, P. Rudecki, M. Becker, S. Schiemann,
M. K\"ulz, and K. Bergmann,
Chem. Phys. Lett. \textbf{149}, 463 (1988).

\bibitem{slowlight}
L. V. Hau, S. E. Harris, Z. Dutton, and C. H. Behroozi,
Nature (London) \textbf{397}, 594 (1999);
M. M. Kash, V. A. Sautenkov, A. S. Zibrov, L. Hollberg,
G. R. Welch, M. D. Lukin, Y. Rostovtsev, E. S. Fry, and
M. O. Scully,
Phys. Rev. Lett. \textbf{82}, 5229 (1999);
D. Budker, D. F. Kimball, S. M. Rochester, and Y. Y. Yashchuk,
{\em ibid.} \textbf{83}, 1767 (1999).

\bibitem{fastlight}
R. Y. Chiao,
Phys. Rev. A \textbf{48}, R34 (1993);
L. J. Wang, A. Kuzmich, and A. Dogariu,
Nature (London) \textbf{406}, 277 (2000).

\bibitem{Yoo85}
H. I. Yoo and J. H. Eberly,
Phys. Rep. \textbf{118}, 239 (1985);
M. O. Scully and M. S. Zubairy,
\textit{Quantum Optics} (Cambridge
University Press, Cambridge, 1997).

\bibitem{spont}
G. S. Agarwal,
\textit{Quantum Statistical Theories of Sponta-neous
Emission and Their Relation to Other Approaches},
Springer Tracts in Modern Physics, Vol. 70,
(Springer, Berlin, 1974);
P. L. Knight,
J. Phys. B \textbf{12}, 3297 (1979);
D. A. Cardimona, M. G. Raymer, and C. R. Stroud,
J. Phys. B \textbf{15}, 55 (1982);
D. J. Gauthier, Y. Zhu, and T. W. Mossberg,
Phys. Rev. Lett. \textbf{66}, 2460 (1991);
J. Javanainen, Europhys. Lett. 17 1992 407;
G. C. Hegerfeldt and M. B. Plenio,
Phys. Rev. A \textbf{46}, 373 (1992);
Quantum Opt. \textbf{6 }, 15 (1994);
H.-R. Xia, C.-Y. Ye and S.-Y. Zhu,
Phys. Rev. Lett. \textbf{77}, 1032 (1996);
G. S. Agarwal,
Phys. Rev. A \textbf{55 }, 2457 (1997).

\bibitem{equivlim}
A. Imamoglu and S. E. Harris,
Opt. Lett. \textbf{14}, 1344 (1989);
M. Fleischhauer, C. H. Keitel, L. M. Narducci,
M. O. Scully, S.-Y. Zhu, and M. S. Zubairy,
Opt. Commun. 94, 599 (1992);
P. R. Berman,
Phys. Rev. A \textbf{58}, 4886 (1999).

\bibitem{Ple00}
M. B. Plenio,
Phys. Rev. A 62, 015802 (2000).

\bibitem{defalg}
M. Rocek,
Phys. Lett. B \textbf{255}, 554 (1991);
D. Bonatsos, C. Daskaloyannis, and G. A. Lalazissis,
Phys. Rev. A \textbf{47}, 3448 (1993);
V. P. Karassiov,
J.Phys. A \textbf{27}, 153 (1994);
V. P. Karassiov and A. B. Klimov,
Phys.Lett. A \textbf{189}, 43 (1994);
C. Quesne,
\emph{ibid} \textbf{193}, 245 (1994),
J. Phys. A \textbf{28}, 2847 (1995);
B. Abdessalam, J. Beckers, A. Chakrabart, and N. Debergh,
J. Phys. A \textbf{29}, 3075 (1996);
N. Debergh,
\emph{ibid} \textbf{30}, 5239 (1997).

\bibitem{Bar77}
A. O. Barut and R. R\c{a}czka,
\textit{Theory of Group Representations and
Applications} (PWN-Polish Scientific, Warszaw,
1977).

\bibitem{Kli00}
A. B. Klimov and L. L. Sanchez-Soto,
Phys. Rev.A, \textbf{61}, 063802 (2000).

\bibitem{Kli02}
A. B. Klimov, J. A. Navarro,
L. L. Sanchez-Soto, and E. C. Yustas,
J.Mod.Opt. \textbf{49}, 2211 (2002).

\end{thebibliography}
\end{document}